-

# Topotactic synthesis, structure and magnetic properties of a new hexagonal polytype of silver cobaltate(III) AgCoO$_{2+\delta}$

Hervé Muguerra[a], Claire Colin[a], Michel Anne[a], Marc-Henri Julien[b], Pierre Strobel[a]

[a]Institut Néel, CNRS and Université Joseph Fourier, B.P. 166, 38042 Grenoble, Cedex 9, France

[b]Laboratoire de Spectrométrie Physique, Université Joseph Fourier, 38402 Saint Martin d'Hères, France

## ABSTRACT

A new form of delafossite-type AgCoO$_{2+\delta}$ was prepared using ion exchange from Na$_{0.75}$CoO$_2$ in molten AgNO$_3$-NH$_4$NO$_3$ at 175°C. Its structure was determined by Rietveld refinement from XRD data; it is hexagonal, space group P6$_3$/mmc, a = 2.871 and c = 12.222 Å. Its structure differs from previously reported AgCoO$_2$ (R-3m, 3R polytype) by the stacking of Co-O layers; in the new phase the 2H stacking of the precursor Na$_{0.75}$CoO$_2$, consistent with a topotactic ion exchange of Na by Ag. The new phase is found to contain a slight oxygen excess ($\delta$ = 0.06). Magnetic susceptibility measurements show the absence of magnetic transition and a weak Curie term, consistent with the non-magnetic character of Co$^{3+}$ ions.

## 1. Introduction

Metal oxides with lamellar structure have attracted wide attention for the past two decades. $LiCoO_2$ is one of the most important intercalation compounds for secondary lithium ion batteries [1]. $Na_xCoO_2$ is recognized to be a potential candidate for thermoelectric materials. More intriguing, superconductivity has been found recently in a hydrated compound of this family: $Na_{0.35}CoO_2,1.3H_2O$ [2].

$LiCoO_2$ and $Na_xCoO_2$ oxides belong to a family of alkali cobalt bronzes ($A_xCoO_2$) with a layered structure, formed of an alternative stacking of $CoO_2$ layers and A planes along the c-axis. The $CoO_2$ layers ($CdI_2$ type) consist of $CoO_6$ distorted octahedra sharing common edges and forming a triangular Co-O lattice. The geometry of this layer is very similar in term of Co-O distances and O-Co-O angles to all of the cobalt thermoelectric compounds [3]. The $CoO_6$ octahedron is compressed along the c direction which results in a distortion of the square plane of the octahedron; this plane became a rectangular plane (O-O : 2.80 and 2.56 Å). Electrical properties of hexagonal phases are of metallic nature and the charge carriers in the system are holes.

In this structure, Na and Co can be substituted by various elements. Cushing and Wiley developed a new method of replacing monovalent ions with divalent ones using a low-temperature ion-exchange technique [4]. They successfully obtained completely substituted $Ca_{0.26}CoO_2$ and $Ca_{0.35}CoO_2$ from $Na_{0.52}CoO_2$ and $Na_{0.75}CoO_2$ precursors respectively. The same technique was used by Ishikawa at al. to obtain $Sr_{0.35}CoO_2$ from $Na_{0.75}CoO_2$ precursor [5]. In all cases, both precursors and their calcium and strontium analogues have similar framework structures. The replacement of the monovalent by divalent cation modifies the electrical properties [5,6]. Some other low temperature synthesis methods exist; $Cs_{0.2}CoO_2,0.63H_2O$ has been obtained from $Na_{0.35}CoO_2,1.3H_2O$ precursor in a CsOH molten flux for example [7]. These different low temperature methods prove the ease of substitution in the sodium layer.

Silver could be an interesting candidate in order to replace Na, since the ionic radii are close ($r[Na^+] = 1.02$Å, $r[Ag^+] = 1.15$Å [8]) and they are both monovalent. Some studies reported Co-Ag exchange by a polymerized complex method [9]. But the resulting products are polyphasic ($Na_{0.75}CoO_2$, $Na_2O_2$ and $Ag_2O$) and no important modifications of the structure or of the physical properties are observed; the diffraction pattern remained practically unchanged [10]. On the other hand, Shin et al. succeeded in preparing $AgNiO_2$ (pure and cobalt-substituted) by ion exchange in silver molten salts [11,12]. $AgCoO_2$ has been known for decades; it has been prepared either from binary oxides in NaOH medium [13,14] or more recently by topotactic exchange [15]; all these studies reported a delafossite-type phase with rhombohedral unit cell. In the present work, we show that ion exchange of sodium by silver in $Na_{0.75}CoO_2$ is a topotactic reaction leading to a new form of $AgCoO_2$ with the same hexagonal cell as the sodium parent compound. This paper reports the synthesis, structure and magnetic behaviour of this new phase.

## 2. Experimental

The Ag compound was prepared in two steps. First $Na_{0.75}CoO_2$ was obtained by solid state reaction between reagent grade $NaHCO_3$ and $Co_3O_4$. An excess of 5 % Na is used to compensate for volatility of sodium carbonate in the temperature range used. The carefully ground mixture was heated 15 hours at 800°C under oxygen flow in an alumina crucible. It was then reground, pelletized, and reheated in same conditions for ca. 48 hours. Secondly, the resulting powder was mixed with $NH_4NO_3$ and $AgNO_3$, and heated under stirring for 15 hours in a temperature range corresponding to the liquid phase in the $NH_4NO_3$-$AgNO_3$ phase diagram [16]. The $NH_4NO_3/AgNO_3$ ratio was chosen to lie in the region near the eutectic composition; exact conditions are summarized in Table 1. A 10 times excess of $AgNO_3$ with respect to $Na_{0.75}CoO_2$ was used in order to promote the exchange $Na^+$-$Ag^+$. The water-soluble byproducts ($NaNO_3$, excess $AgNO_3$, $NH_4NO_3$) were removed by washing in distilled water. X-ray powder diffraction measurements (XRD) were carried out using a Brucker D8 Advance diffractometer equipped with $Cu_\alpha$ source. The diffraction intensities were measured in the range from $2\theta = 10°$ to $100°$ at a step width of 0.025. The crystal structure of the new $AgCoO_{2+\delta}$ was refined by the Rietveld analysis program FullProf software suite [17]. Theermogravimetric experiments were performed with a Setaram TAG 1750, using a heating rate of 0.2°C/min under argon and in oxygen.

The magnetic properties were measured with a superconducting quantum interference device (SQUID) magnetometer in the range 4–300 K. M(T) data were recorded in a magnetic field of 100Oe.

## 3. Results and discussion

### 3.1 Synthesis

Different experimental conditions were tried in order to obtain the AgCoO$_{2+\delta}$ phase without impurities and in the best crystallised form. The reaction was carried out at 200°C (run 1) and 175°C (runs 2 to 4, see Table 1). The latter temperature gave the more conclusive results; indeed 200°C is close to the decomposition temperature of NH$_4$NO$_3$ (210°C). At this temperature, we noted that after one night the reaction medium had solidified, indicating a modification in the composition mixture (evaporation and/or decomposition NH$_4$NO$_3$). Moreover the Ag/Co ratio was inhomogeneous in the solid product of run 1, while it was constant (Ag/Co≈1/1) in runs carried out at 175°C. A temperature of 175°C was then chosen because it is high enough to favour cation mobility and exchange without leading to molten salt decomposition.

First attempts yielded an AgCoO$_{2+\delta}$ phase containing Co$_3$O$_4$ impurity (< 1 wt.%), whereas it was absent in the Na$_{0.75}$CoO$_2$ precursor. A change in the NH$_4$NO$_3$/AgNO$_3$ molar ratio (3/2, run 3) did not yield any improvement. However, we noticed that the molten reaction mixture at 175°C contained a small fraction of solid phase at the bottom of the crucible. In run 4, this solid was separated from the liquid phase at high temperature. After cooling and washing the solidified liquid in distilled water, it was found that the precipitate did not contain any trace of cobalt oxide. This shows that during the reaction at 175°C, fine-grained Na$_{0.75}$CoO$_2$ remains in suspension in the liquid whereas Co$_3$O$_4$ falls at the bottom of the reaction vessel and can therefore be separated.

The elemental analysis performed by energy X-ray dispersive spectroscopy on the compound formed gave a cationic ratio Ag/Co = 49/51, with a negligible presence of sodium.

Tests of intercalating water molecules in the new phase showed that, contrary to the Na$_{0.75}$CoO$_2$ precursor, new AgCoO$_{2+\delta}$ is not moisture-sensitive.

### 3.2 Phase identification

Figure 1 shows the diffraction pattern of the phase obtained, compared to that of the known form of AgCoO$_2$, clearly showing important differences. In fact, all observed reflections of the new phase can be indexed in a cell isotypic to that of the precursor used, Na$_{0.75}$CoO$_2$ (space group P6$_3$/mmc). The unit cell, however, is significantly larger, with a = 2.871, c = 12.222 Å, compared to 2.840 and 10.811Å for Na$_{0.75}$CoO$_2$. The increase in cell volume can be ascribed to the difference in ionic radius between Ag$^+$ and Na$^+$ (r[Ag$^+$] = 1.15 and r[Na$^+$] = 1.02 Å [8]). It seems therefore that we obtained a new form of AgCoO$_{2+\delta}$, differing from the previously reported one by the stacking along c (solid-state route AgCoO$_2$: R-3m, a = 2.873 and c = 18.336Å). This new phase will be abbreviated in the following "AgCoO$_{2+\delta}$ (IE)" (for Ion-Exchanged). Interestingly, the phase obtained here differs from that obtained using the same preparation route in molten salt by Shin et al. [12], perhaps due to the fact that their study concerned nickel-containing phases $AgNi_{1-x}Co_xO_2$.

### 3.3 Structure determination and description

The $Na_{0.75}CoO_2$ structure (space group $P6_3/mmc$) was considered as initial structural model for the refinement. It was quickly realized that the interlayer space contained only one Ag site ($^1/_3;^2/_3;^1/_4$) instead of two Na sites ($0;0;^1/_4$ and $^2/_3;^1/_3;^1/_4$) in the parent sodium cobaltate. The positions of cobalt ($0;0;^1/_2$) and oxygen ($^1/_3;^2/_3;z$), are found unchanged, with a slight shift in the z coordinate [from z =0.0913(1) to z=0.076(1)]. The existence of vacancies on all atoms sites was probed and found negligible.

Using March's function, a [001] preferred orientation was found and included in the refinement. The background was interpolated from 60 experimental points. Final variables were the scale factor, zero shift, cell parameters, profile parameters, oxygen z-coordinate, isotropic displacement parameters B for all atoms, and preferred orientation. The results of refinement are summarized in Table 2. Figure 2 shows observed, calculated and difference profiles. The B factor of the silver atom appears to be higher than those of cobalt and oxygen atoms. This feature is consistent with weaker Ag-O bonds and significant cation mobility in the interlayer space.

The replacement of sodium by silver in the hexagonal framework brings an important change, namely the coordination of the interlayer cation, which decreases from 6 in $Na_{0.75}CoO_2$ to 2 in $AgCoO_{2+\delta}$ (IE). The linear O-Ag-O coordination parallel to the hexagonal c-axis is typical of the delafossite structure (Fig.3). However, several stackings of transition metal-oxygen layers are possible within this structural family (Fig. 3). If successive O layers are labelled O1, O2 and O3, possible sequences are:

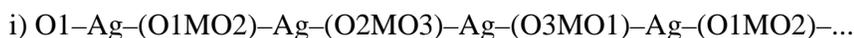

i) O1–Ag–(O1MO2)–Ag–(O2MO3)–Ag–(O3MO1)–Ag–(O1MO2)–...

with two successive shifts of the oxygen layers in the same direction (A–B–C–A–B–C... stacking). This yields a rhombohedral structure with a 3-layer periodicity along z ("3R" structure) : c ≈ 18 Å.

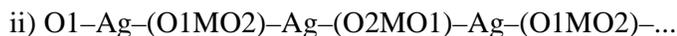

ii) O1–Ag–(O1MO2)–Ag–(O2MO1)–Ag–(O1MO2)–...

with the third oxygen layer at the same *x,y* position as the first one (A–B–A–B... stacking). This yields a hexagonal structure with a 2-layer periodicity along z ("2H" structure) : c ≈ 12 Å.

In this study, the use of a *chimie douce* route from a 2H precursor allowed to maintain the A–B–A–B oxygen stacking and to obtain a new 2H-delafossite polytype instead of the more common 3R one. This difference in stacking is clearly reflected in the space group ($P6_3/mmc$ instead of R-3m) and c-cell parameter (12.222 instead of 18.336Å).

A survey of known silver-transition metal $AgMO_2$ delafossites shows the existence of 3R or 2H polytypes depending on the transition element M (see Table 3) : the 2H form was found for M = Fe, whereas chromium forms a 3R delafossite. Interestingly, both forms have been reported in the Ag-Ni-O [11,18].

A comparison of Co-O and O-O distances after and before Ag ion exchange shows only minor modifications in the cobalt-oxygen layer (Tab.3): the Co-O distances shifted from 1.914 to 1.902Å after Na-Ag exchange. The $CoO_6$ octahedra are compressed; and their distortion is conserved (O-O 2.871 and 2.495Å, O-Co-O 98° and 82° in $AgCoO_{2+\delta}$ (IE)]. Note that this feature is common to $Na_xCoO_2$ and other delafossite-type oxides (see O-O distances and O-Co-O angles in Table 3). The Ag-O distances are quite similar ($\approx 2.10$ Å) in the $AgMO_2$ delafossite family (M = Ni, Co, Fe, Cr), Ag-Ag distances are close to 2.9 Å, which is comparable to the Ag-Ag distance in metallic silver, 2.889Å (Tab.3).

### 3.4. Oxygen and cation stoichiometry

The calculated bond valence of Co atoms is +3.47 using Brese and O'Keeffe parameter [22], or +3.36 with reference to the $Li_xCoO_2$ system [23]. The latter value seems more plausible, because of the similarity of $AgCoO_2$ and $Li_xCoO_2$ structures. But all these values differ notably from the formal cobalt valence (+3). This could be explained by oxygen non-stoichiometry ($\delta > 0$ in the formula $AgCoO_{2+\delta}$). Oxygen excess in delafossite structures was reported in the $CuMO_2$ family (M = Y, La, Pr, etc.), where several groups [24,25] have demonstrated the possibility to insert oxygen atoms in the Cu layers, namely in the center of Cu triangles. Very recently, oxygen excess in other silver delafossites was also reported [26]. In order to determine more precisely the oxygen stoichiometry, TGA analyses were carried out in oxygen and in argon. In the first case, no significant weight change was observed up to 600℃. In argon, the thermogravimetric measurement up to 750℃ (Fig.4) showed a slight mass loss $\Delta m_1$ in the range 100-575℃ followed by an abrupt one $\Delta m_2$ above 575℃. The mass is constant (i.e. stable compounds are formed) above 660℃. The XRD analysis of the final product shows the presence of metallic silver and $Co_3O_4$, so that the overall reaction in argon can be written as follows:

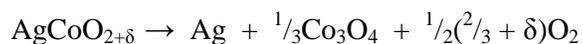

$$AgCoO_{2+\delta} \rightarrow Ag + {}^1\!/_3Co_3O_4 + {}^1\!/_2({}^2\!/_3 + \delta)O_2$$

The δ value calculated from the mass loss assuming this reaction yields δ = +0.058 ± 0.005. Moreover, the mass losses $\Delta m_1$ and $\Delta m_2$ are consistent with the following, separate reactions:

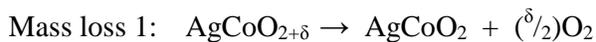

Mass loss 1: $AgCoO_{2+\delta} \rightarrow AgCoO_2 + (^\delta/_2)O_2$

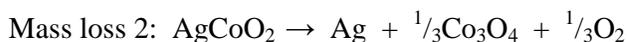

Mass loss 2: $AgCoO_2 \rightarrow Ag + {}^1\!/_3 Co_3O_4 + {}^1\!/_3 O_2$

This experiment confirms the presence of oxygen over-stoichiometry in the new hexagonal Ag-Co-O phase. This extra oxygen is gradually lost in the temperature range 100-575°C, at which point the ternary oxide is destroyed.

The value of δ obtained by this measurement (0.058) is much lower than the one obtained from bond valence calculations ($Co^{+3.36}$ yielding δ = 0.18). Note that the bond valence calculation requires bond length standards, which are available for the $Co^{+3}$-O bond, but not for the $Co^{+4}$-O one; this may result in a poor reliability of calculated Co bond valence for Co > 3+. Finally, attempts to locate the extra oxygen atoms from XRD structure refinements were unconclusive, because of the light weight of the oxygen atom and the presence of heavy silver atoms in the same layer perpendicular to the c axis.

It is worth pointing out that, in spite of the fact that all previous preparations of $AgCoO_2$ yielded the 3R form, the 2H form is stable at least up to 750°C. XRD taken after the thermogravimetric experiment (maximum temperature 750°C) indeed show that the new compound is not transformed into the 3R form at this temperature. The difference in thermodynamic stability between the 3R and 2H forms is likely to be small, as shown in the Cu-Y-O system where mixtures of both stackings are usually obtained [27].

Regarding cation stoichiometry, the parent phase $Na_xCoO_2$ is well known for accommodating a variable concentration of sodium vacancies. One of its most remarkable properties, the occurrence of superconductivity at 4-5 K, is obtained after partial sodium extraction coupled with water molecules intercalation [2]. Attempts to perform similar oxidative de-intercalation of silver with bromine or permanganate were unsuccessful. The increased stability of the interlayer cation slab can be ascribed to the delafossite structure, where silver atoms form fixed pillars between cobalt-oxygen layers; this situation differs considerably from the interlayer configuration in $Na_xCoO_2$, where sodium atoms are shared between two different sites which can easily accept partial occupation. This is also the probable reason why the overall cation occupation changes during the sodium-silver ion exchange, resulting in reduction of cobalt oxidation state from +3.3 to +3.

### 3.5. Magnetic properties

Figure 5 shows the magnetic susceptibility $\chi$ as a function of temperature, measured in a field of 100 Oe. No significant difference was found between field-cooled and zero-field-cooled measurements. The data are well described by the sum of a constant term $\chi_0$ mostly due to orbital (Van Vleck) paramagnetism and of a small Curie term $C/T$. A fit of the experimental data with the equation

$$\chi = \chi_0 + C/(T+\Theta)$$

gives $\chi_0 = 6.0 \ 10^{-5}$ emu mol$^{-1}$, $C = 9.5 \ 10^{-4}$ emu K mol$^{-1}$ and $\Theta \approx 0$. The $\chi_0$ value is consistent with the cobalt Van Vleck paramagnetic susceptibility. The Curie term corresponds to 0.087 $\mu_B$/formula unit, i.e. 0.05 Co$^{4+}$, in fairly good agreement with oxygen over-stoichiometry. The absence of interaction between these magnetic moments is demonstrated by the fact that the Curie-Weiss temperature, if forced into the fit, is zero within experimental accuracy. The weakness of the magnetism, and particularly the absence of any magnetic transition at low temperature, are consistent with the S=0 spin ground state of Co$^{3+}$ ions in A$_x$CoO$_2$ materials, which are non magnetic band insulators for x=1 [28,29]. Such results are consistent with weak paramagnetism observed previously on 3R-AgCoO$_2$ by Shin et al. [12]. Note that non-zero spins have also been frequently reported in LiCoO$_2$; they were attributed to a possible intermediate spin state for cobalt [30]; this is rather unlikely here since all Co–O interatomic distances are equal.

## 4. Conclusion

A new AgCoO$_{2+\delta}$ phase has been obtained by a topotactic, *chimie douce* reaction at 175°C. This phase differs from previously reported AgCoO$_2$ by the stacking of Co-O layers, which is reflected in the space group (P6$_3$/mmc instead of R-3m) and the c cell (12.222 instead of 18.336Å). This new cobalt oxide has the main stacking than the Na$_{0.75}$CoO$_2$ precursor used in its synthesis. However, the structural refinement showed that the interlayer cation occupation is different: the two Na sites of the precursor are replaced by one Ag site forming linear AgO$_2^{3-}$ groups parallel to the hexagonal c-axis. In this work, we also show that this new phase contains a slight oxygen excess, as has been found in a number of copper delafossites. Magnetic measurements showed a non-zero magnetic moment and the absence of long-range ordering down to 4 K. This work is a new example of the interest of ion exchange in eutectic molten salts at low temperature. This synthesis method is a very valuable route to prepare new forms of layered materials, in some cases metastable, by ion exchange. Extensions to other layered cobaltates are certainly to be considered.

**Acknowledgments**

This work was supported by the French Agence Nationale de la Recherche (Grant No. 06-BLAN-0111). The authors also thank X. Chaud for his assistance in thermogravimetry measurements and O. Isnard for fruitful discussions.

**Table 1:** Synthesis conditions for ion exchange of Ag in $Na_{0.75}CoO_2$.

| Run No | $NH_4NO_3/AgNO_3$ ratio | T (°C) | Time | Notes |
|---|---|---|---|---|
| 1 | 2/3 | 200 | 15h | solid after 15 h |
| 2 | 2/3 | 175 | 15h | contains $Co_3O_4$ impurity |
| 3 | 3/2 | 175 | 15h | ibid. |
| 4 | 2/3 | 175 | 15h | $Co_3O_4$ impurity separated from liquid phase at 175°C |

**Table 2:** Results of structure refinement for $AgCoO_{2+\delta}$ (IE) at 293 K.

| Atom | Site | x | y | z | Occupancy | B (Å$^2$) |
|---|---|---|---|---|---|---|
| Ag | 2c | $^1/_3$ | $^2/_3$ | $^1/_4$ | 1 | 1.473(7) |
| Co | 2a | 0 | 0 | 0 | 1 | 1.171(9) |
| O | 4f | $^1/_3$ | $^2/_3$ | 0.076(1) | 1 | 1.157(3) |

| | | |
|---|---|---|
| Lattice Parameters | a (Å) | 2.87091(5) |
| | c (Å) | 12.22159(7) |
| Space group | $P6_3/mmc$ | |
| | $R_{wp}$ (%) | 4.58 |
| R factors | $R_{Bragg}$ (%) | 5.54 |
| | $\chi^2$ | 4.47 |
| Preferred orientation | [001] | 1.65537 |

**Table 3:** Structural characteristics, selected bond lengths and angles for the $Na_{0.75}CoO_2$ and different silver delafossite compounds. *SG = Space Group.*

| $AMO_2$ | a (Å) | c (Å) | SG | M-O (Å) | O-O (Å) | O-M-O (°) | Ag-O (Å) | Ag-Ag (Å) |
|---|---|---|---|---|---|---|---|---|
| $Na_{0.75}CoO_2$ [19] | 2.840 | 10.811 | $P6_3/mmc$ | 1.914 | 2.840 / 2.566 | 95.8 / 84.2 | - | - |
| $AgCoO_{2+\delta}$ (IE) [Present study] | 2.871 | 12.222 | $P6_3/mmc$ | 1.902 | 2.871 / 2.495 | 98.0 / 82.0 | 2.123 | 2.871 |
| $AgNiO_2$ [18] | 2.937 | 12.237 | $P6_3/mmc$ | 1.954 | 2.937 / 2.579 | 97.4 / 82.6 | 2.088 | 2.937 |
| $AgFeO_2$ [20] | 3.039 | 12.395 | $P6_3/mmc$ | 2.036 | 3.039 / 2.710 | 96.6 / 83.5 | 2.066 | 3.039 |
| $AgCoO_2$ [14] | 2.873 | 18.336 | R-3m | - | - | - | - | 2.875 |
| $AgNiO_2$ [11] | 2.939 | 18.370 | R-3m | 1.940 | 2.939 / 2.532 | 98.5 / 81.5 | 2.122 | 2.939 |
| $AgCrO_2$ [21] | 2.985 | 18.510 | R-3m | 1.999 | 2.985 / 2.661 | 96.6 / 83.5 | 2.071 | 2.985 |

- 13 -

**FIGURE CAPTIONS**

**Figure 1:** Comparison between the XRD patterns of $AgCoO_{2+\delta}$ (IE) phase and rhombohedral $AgCoO_2$ . The latter is a simulated diagram using Shannon's structural data and the same preferred orientation as $AgCoO_2$(IE).

**Figure 2:** Observed (*dot*) and calculated (*solid line*) powder X-ray diffraction intensities of the $AgCoO_{2+\delta}$ (IE) compound. Tick marks indicate the positions of Bragg reflections. A solid line at the bottom shows the difference between observed and calculated intensities.

**Figure 3:** (a) 3R-and (b) 2H-delafossite-type structures of $AMO_2$ oxides.

**Figure 4:** Thermogravimetric curve of $AgCoO_{2+\delta}$ (IE) under flowing argon.

**Figure 5:** Magnetic susceptibility of $AgCoO_{2+\delta}$(IE) as a function of temperature (T) measured in Zero Field-Cooling (ZFC) mode with H = 100Oe.

**N ÉEL**
institut

| | |
|---|---|
| **Pierre STROBEL**<br>*Directeur de Recherches* | TEL : + 33 (0)476 887 940<br>FAX : +33 (0)476 881 038<br>*pierre.strobel@grenoble.cnrs.fr* |

### *Response to Reviews*

### *manuscript JSSC-08-771*

We thank the reviewers for numerous interesting remarks. They were taken into account as follows:

Reviewer #1: the question of silver non-stoichiometry and cobalt oxidation state is now addressed in section 3.4 (one additional paragraph).

Reviewer #2:

1. This point is now addressed in section 3.2; our study differs from that in ref.12 because they actually studied the $AgNi_{1-x}Co_xO_2$ system; on the other hand Ref. 14 did not use ion exchange.

2. The yield has little significance in this kind of synthesis where a significant portion of the sample is lost in separation and filtration processes. An upper estimate of the $Co_3O_4$ fraction has been included in the text.

3. We thank the referee for this useful reference, which has been included (ref.26 in new version).

4. The printout of the EDX analysis has been included as supplementary information.

Reviewer #3:

1. This is a comment; earlier works are cited (refs.11 and 12)

2. This reference has been added in the introduction (new ref.15)

3. This is a comment connected to point 6 - see below

(no point 4)

5. The abbreviation "6H" has been corrected to "2H" throughout the text and figures.

6. Stability of the 2H vs. 3R form: this point is now addressed in section 3.4.

7. We do not believe that such small oxygen non-stoichiometry can play a role; no previous reference gives any indication of this.

8. We explain this possibility of separation by the difference in grain size between layered $A_xCoO_2$ and $Co_3O_4$; this is now commented in the text, section 3.1.

9. The caption of Fig.5 has been correeted.

10-12. The units for $\chi$ and C in the text and in fig.5 have been corrected

13. The Van Vleck term is consistent with that expected for cobalt; this is now included in the text, section 3.5.

14. Typing errors corrected.

15. See remark of reviewer 1.

16 . We agree with this comment. It is mentioned at the end of section 3.5.

Minor spelling errors were corrected, and the references were renumbered.

We hope that this manuscript is now acceptable for publication.

Pierre STROBEL

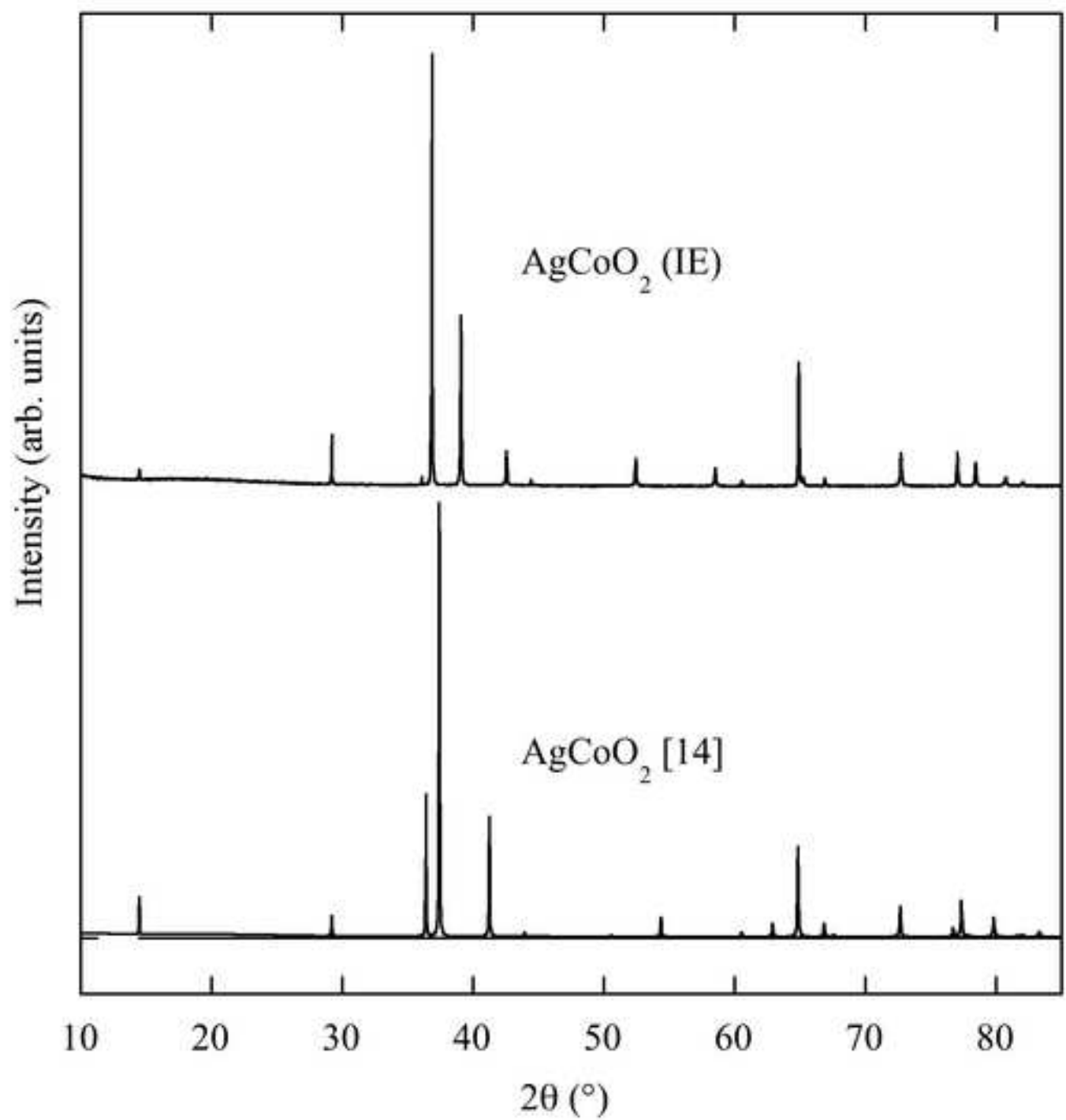

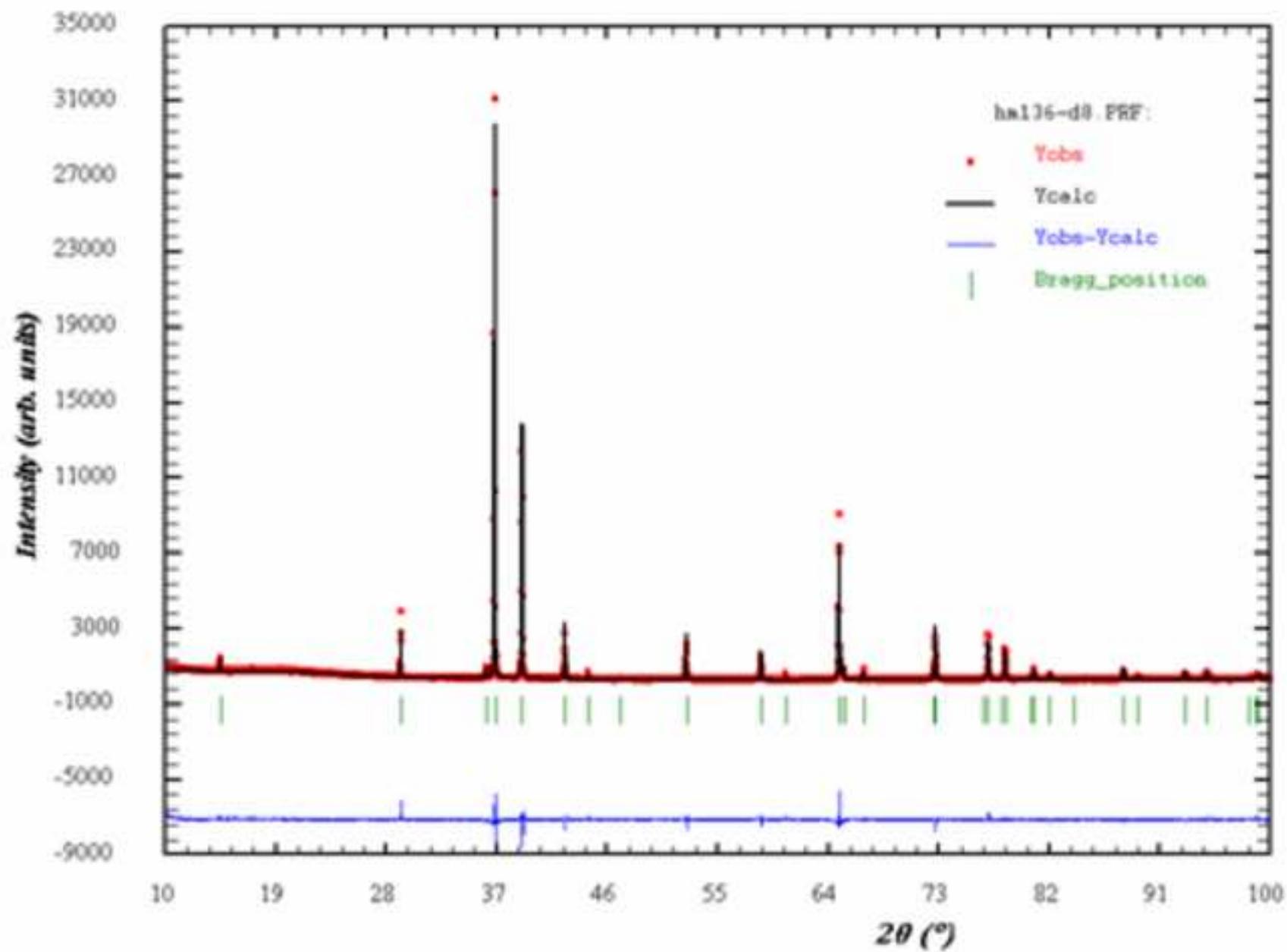

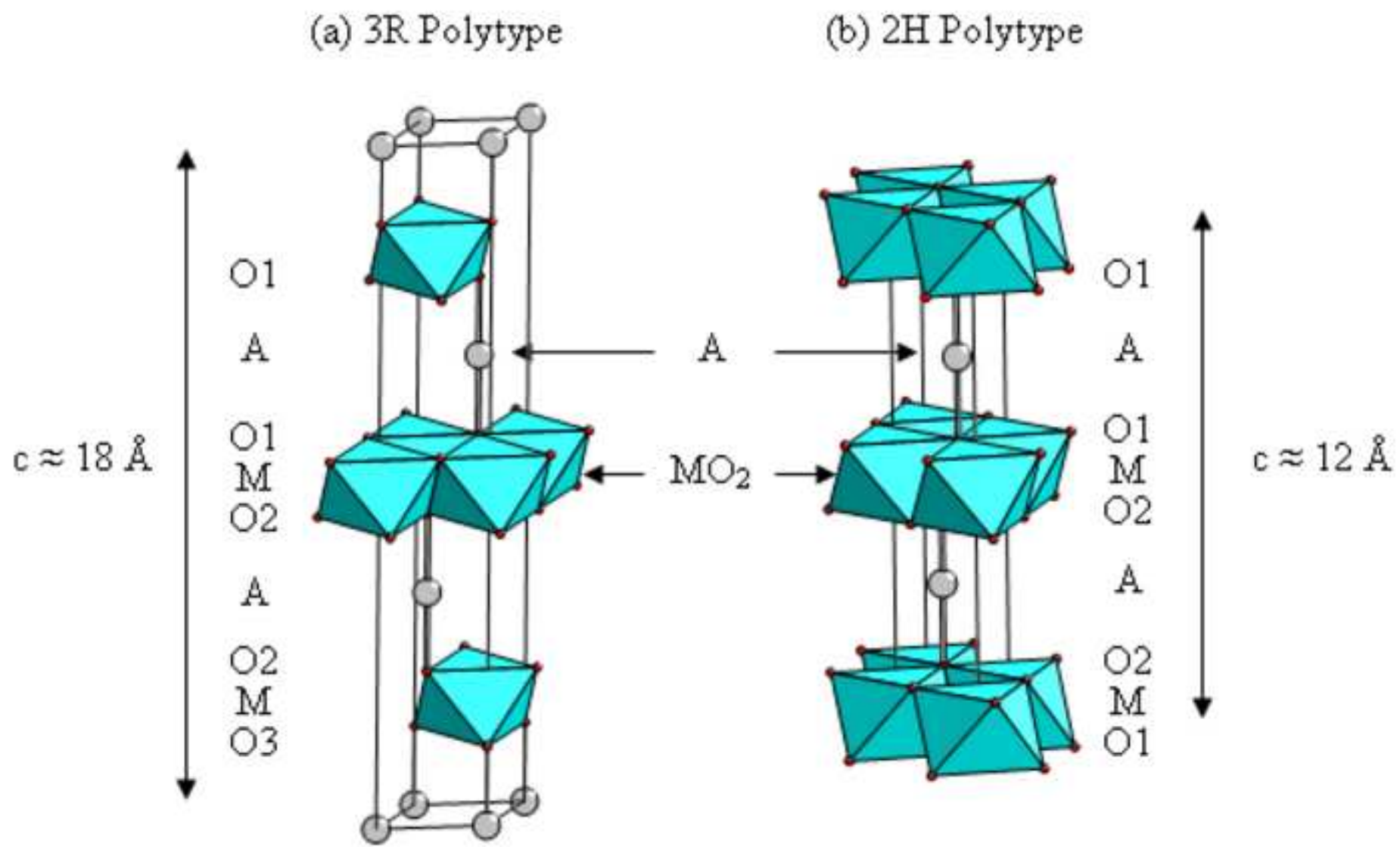

(a) 3R Polytype        (b) 2H Polytype

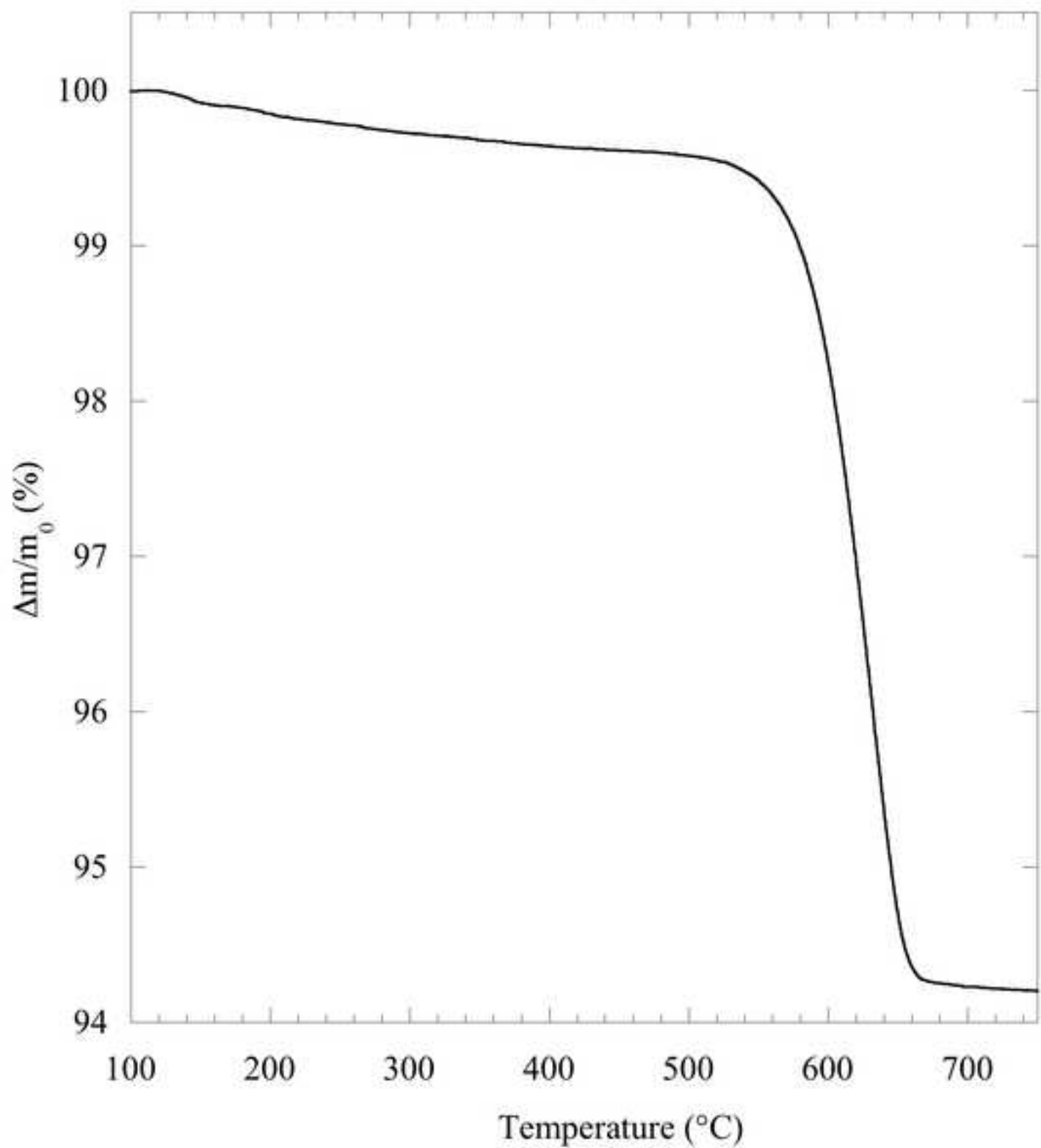

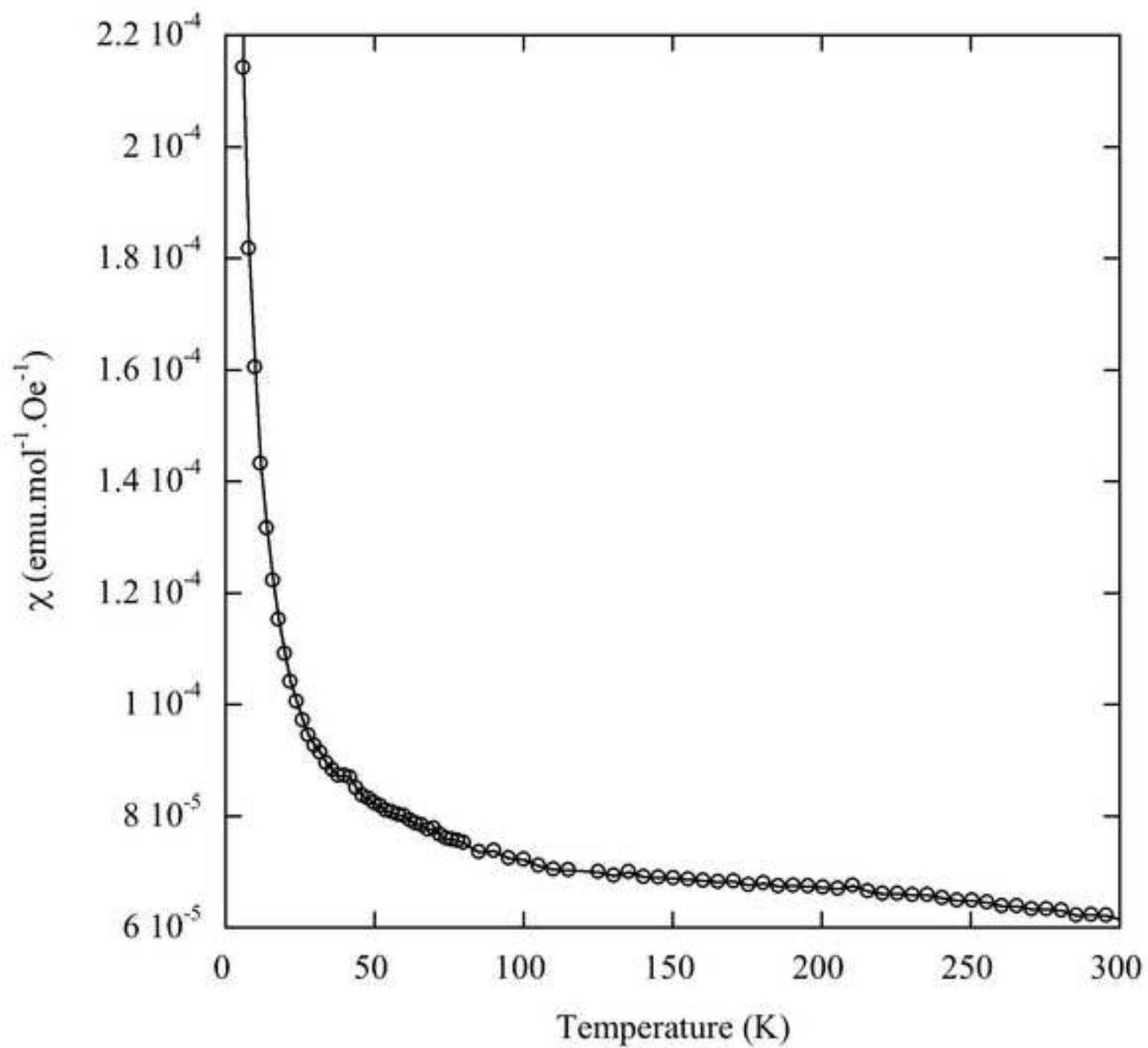

**Supplemental Data**

Click here to download Supplemental Data: 08-771_annex_EDX.TIF